\documentstyle[12pt]{article}
\input tcilatex
\begin{document}

\begin{center}
{\bf The three dimensional theory of gravitation based on a temporal scalar
field in Riemanian geometry.}

\vspace*{1.5cm} S.M.KOZYREV

e-mail: Sergey@tnpko.ru
\end{center}

\vspace*{1.5cm}

\begin{center}
${\bf Abstract}$
\end{center}

A new approach to the model of the universe based on work by Rippl, Romero,
Tavakol is presented. We have used the scheme for relating the vacuum (D +
1) dimensional theories to D dimensional theories for setting up a
correspondence between vacuum 4-dimensional Einstein theory with
3-dimensional gravity theory with temporal scalar field. These ideas we
continued by using the 3-dimensional analog of Jordan, Brans-Dicke theory
with temporal scalar field. As the result space and time are treated in
completely different ways. For the case of a static spherically symmetric
field new vacuum static solutions are found.

\section{Introduction}

The problem of time in relativistic theories of gravity is deeply connected
with the special role assigned to temporal concepts in standard theories of
physics. In particular, in Newtonian physics, starting point is a
three-dimensional manifold which serves as a model for physical space and
time is external to the system itself. This is contrasted with the so-called
'spacetime' (or 'covariant') approaches in which the basic entity is a
four-dimensional spacetime manifold. Since the advent of general relativity,
the aim of a purely geometrical description of all physical forces as well
as that of geometrical origin of the matter content of the Universe has been
pursued \cite{1}.

Multidimensional purely geometrical gravity has been of much interest at
least since the pioneering ideas of Kaluza \cite{10} and Klein \cite{2}.
These ideas were continued by Jordan \cite{20} who suggested considering the
more general case g$_{55}\neq $ const leading to the theory with an
additional scalar field. It is useful here to emphasise a different
interpretation of this scheme by recalling that for 5-dimensional
Kaluza-Klein spacetimes with a symmetry group generated by (implying that
the metric is independent of the extra spatial dimension) the field
equations are formally identical to the vacuum Jordan, Brans-Dicke field
equations in four dimensions with the free parameter $\omega $ = 0.

Scalar fields are today again being investigated and with greater vigor than
ever before. Recently \cite{3} there has been some interest in the idea that
the time is a scalar field on the manifold. Motivated by Kaluza-Klein
reduction procedure from (D+1) dimension one can studied a spacetime
containing temporal scalar field. These ideas can be extended to systems
that are compatible with Newtonian physics: one simply replaces the time
system of with the set of temporal scalar field. The addition of the
temporal scalar field to the Einstein tensor gives a unique theory. Jordan
and his colleagues took the step of separating the scalar field from the
original five-dimensional metric context unified
gravitational-electromagnetic context having applied this schema one can
separate the temporal scalar field from the four-dimensional metric context.

This correspondence between the vacuum Einstein theory and lower dimensional
theories is of potential importance in view of the radical differences
between such theories and 4-dimensional gravity. Space and time are treated
in completely different ways. This might seem to be a retrogressive step,
but actually the approach achieves everything that Einstein did by
presupposing a four-dimensional unity of space and time. It is important to
note in this connection that there is a fundamental difference between
physics and geometry. In geometry there is no motion that is tightly
connected with the concept of time.

A basic idea for attempting to formulate a theory of gravitation which is
satisfactory from the standpoint of temporal scalar field is considered in
Sec. 2. In Sec. 3 the exact static spherically symmetric vacuum solutions to
these theory has been obtained. A similar solution has been carried out in
Sec. 4 for 3-dimensional more generalized theory with $\omega $ $\neq $ 0.
Finally, Sec. 5 briefly discussed some aspects of these theories. Convention
in this paper follow those of \cite{c31} with Newtonian constant G =1 and c
= 1. Here, Greek indices take on values 1 $\rightarrow $ D whereas Latin
indices take on values 1 $\rightarrow $ D+1.

\section{Basic Ideas.}

Consider the case of (D + 1) field vacuum equations

\begin{center}
\begin{equation}
^{D+1}R_{ab}=0,  \label{1.1}
\end{equation}
\end{center}

where R$_{ab}$ is a Ricci tensor in the (D+1) - dimensional space. It is
well known that equation (\ref{1.1}) gives rise to D dimensional Einstein
equations with source in the form \cite{c32}:

\thinspace 
\begin{eqnarray}
^DR_{\alpha \beta }=^DT_{\alpha \beta }.  \label{1.2}
\end{eqnarray}

Provided extra terms related to the (D+1) dimension are appropriately used
to define an energy momentum tensor T$_{a\ss }$. Furthermore, the matter
source in the D dimensions for that interpretation can be determined by
using reasonable equation of state \cite{c31}. The investigation of a
similar theory of gravity where (D+1) dimension is a time appears to be the
one of the way to construct a model with vacuum properties akin to those of
general relativity. Although this removes the scenario from Einsteins theory
it is nevertheless of interest to seek a minimal modification that can
accommodate such features as temporal field. We start with the (D +1)
dimensional source-free Kaluza-Klein field equations (\ref{1.1}) and take
the metric to be in the form:

\begin{equation}
\left( 
\begin{array}{cccccc}
&  &  &  &  & 0 \\ 
&  &  &  &  & 0 \\ 
&  & g_{\alpha \beta } &  &  & 0 \\ 
&  &  &  &  & . \\ 
&  &  &  &  & . \\ 
0 & 0 & 0 & . & . & g_{dd}
\end{array}
\right) ,  \label{1.3}
\end{equation}

where d = D + 1 and both the (D + 1) and D dimensional metrics g$_{ab}$ and g%
$_{a\ss }$\ are in general dependent on the coordinate x$^d$ and we write
the g$_{dd}$ as

\begin{equation}
g_{dd}=\phi ^2,g^{dd}=\frac 1{\phi ^2}.  \label{1.4}
\end{equation}

Note that one could obtain part of the Jordan, Brans-Dicke action with the
free parameter $\omega $ = 0 \cite{c33}. Now the source free field equations
in (D + 1) dimensions are given by

\begin{eqnarray}
\begin{array}{cc}
& ^{(D+1)}R_{\alpha \beta }=0 \\ 
^{(D+1)}R_{ab}=0\Rightarrow & ^{(D+1)}R_{dd}=0 \\ 
& ^{(D+1)}R_{\alpha d}=0
\end{array}
\label{2.2}
\end{eqnarray}

and

\begin{equation}
\begin{array}{l}
^{(D+1)}R_{\alpha \beta }=\,^{\,(D)}R_{\alpha \beta }- \\ 
\,\,\,\,\,\,\,\,\,\,\,\,\,\,\,\,\,\,\,\,\,\,\,\,\,\,\,-\frac{\phi _{\alpha
;\beta }}\phi \frac 1{2\phi ^2}\left( \frac{\phi ,_d}\phi g_{\alpha \beta
,d}-g_{\alpha \beta ,dd}+g^{\lambda \mu }g_{\alpha \lambda ,d}g_{\beta \mu
,d}-\frac 12g^{\mu \nu }g_{\mu \nu ,d}g_{\alpha \beta ,d}\right) ,
\end{array}
\end{equation}

In this way the (D + 1)-dimensional source free Einstein type equations are
related to a D dimensional theory with sources, from the first of these (\ref
{2.2}) we obtain

\begin{equation}
^{(D)}R_{\alpha \beta }-\frac 12\,^{(D)}Rg_{\alpha \beta }=^{(D)}T_{\alpha
\beta ,}  \label{1.5}
\end{equation}

where an effective energy-momentum tensor is

\begin{equation}
\begin{array}{r}
^{(D)}T_{\alpha \beta }=\frac{\phi _{\alpha ;\beta }}\phi -\frac 1{2\phi
^2}\left[ \frac{\phi ,_d}\phi g_{\alpha \beta ,d}-g_{\alpha \beta
,dd}+g^{\lambda \mu }g_{\alpha \lambda ,d}g_{\beta \mu ,d}-\frac 12g^{\mu
\nu }g_{\mu \nu ,d}g_{\alpha \beta ,d}\right] + \\ 
+\frac{g_{\alpha \beta }}{8\phi ^2}\left[ g^{\mu \nu },_{d\,\,}g_{\mu \nu
,d}+\left( g^{\mu \nu }g_{\mu \nu ,d}\right) ^2\right] .
\end{array}
\label{1.6}
\end{equation}

Note that in more general case stress-energy tensor is assumed to have been
derived from the Lagrangian for matter in the usual way. The second wave
equation for the scalar field (\ref{2.2}) obtained in the usual way

\begin{equation}
\ \Box \phi =-\frac 1{4\phi }g^{\lambda \mu }\,_{,d\,}g_{\lambda \nu
,d}-\frac 1{2\phi }g^{\lambda \mu }g_{\lambda \mu ,dd}+\frac{\phi ,_d}{2\phi
^2}g^{\lambda \mu }g_{\lambda \mu ,d},  \label{1.7}
\end{equation}

Here the generally covariant d'Alembertian $\Box $ is defined to be
covariant divergence of $\phi ^{,\rho }$:

\[
\Box \phi =\phi ^{,\rho }\,_{;\rho }=\sqrt{-g}\left( \sqrt{-g}\phi ^{,\rho
}\right) _{,\rho }. 
\]

Finally, the last equations in (\ref{2.2}) have the appearance of
conservation laws, the full meaning of which is not clear at the present.

\begin{equation}
\ \left[ \frac 1{2\sqrt{g_{dd}}}\left( g^{\beta \mu }g_{\mu \alpha
,d}-\delta _\alpha ^\beta \,g^{\rho \nu }g_{\rho \nu ,d}\right) \right]
_{;\beta }.  \label{1.8}
\end{equation}

\section{Static spherically symmetric field.}

As a first step, we employ the 4 dimensional source-free Einstein field
equations (\ref{1.1}) and take the metric to be in the form (\ref{1.3})
where both the 4 and 3 dimensional metrics g$_{ab}$ and g$_{a\ss }$\ are in
general dependent on the coordinate {\it t} and one can write the g$_{44}$ as

\begin{eqnarray}
g_{44}=\phi ^2,g_{^{44}}=\frac 1{\phi ^2}.  \label{2.3}
\end{eqnarray}

In empty space when the energy-momentum tensor of matter vanishes, the
system of equations (\ref{1.5}), (\ref{1.7}) have the vacuum solution g$%
_{\mu \nu }$ = $\eta _{\mu \nu }$ and $\phi $ = const, where $\eta _{\mu \nu
}$ is the Minkowski metric tensor. This solution fixes the arbitrariness in
the coordinate system. Now for the static spherically symmetric field we
consider line element in the form

\begin{eqnarray}
ds^2=e^\lambda dr^2+r^2\left( d\theta ^2+\sin ^2\theta \,d\varphi ^2\right) .
\label{2.5}
\end{eqnarray}

Then the gravitational field equations and equation for the scalar temporal
field yield

\begin{eqnarray}
\begin{array}{c}
\frac{\lambda ^{\prime }}r+\frac{\lambda ^{\prime }\phi ^{\prime }}{2\phi }-%
\frac{\phi ^{\prime \prime }}\phi =0 \\ 
\\ 
2\left( e^{\lambda \,}-1\right) +r\,\lambda ^{\prime }-\frac{2r\,\phi
^{\prime }}\phi =0 \\ 
\\ 
-\frac 4r+\lambda ^{\prime }-\frac{2\,\phi ^{\prime \prime }}\phi =0
\end{array}
\label{2.7}
\end{eqnarray}

where the primes stand for derivation respect to {\it r}. The solution of
these equations is given by

\begin{eqnarray}
\begin{array}{c}
e^\lambda =\frac r{r\,+\,m} \\ 
\\ 
\phi =\sqrt{-1-\frac mr}
\end{array}
,  \label{2.6}
\end{eqnarray}

where {\it m} integration constant. Obviously in general relativity that is
identical to the Schwarzschild solution. We should note, however, that
despite the mathematical equivalence of the two solutions, that fact that
they come from different physical theories makes them conceptually distinct. 

\section{Three dimensional temporal scalar fields theory with $\omega \neq $
0.}

From the above it is seen that the field equations (\ref{1.5}) - (\ref{1.7})
of this model are formally identical to the vacuum Jordan, Brans-Dicke field
equations in three dimension with the free parameter $\omega $ = 0. However,
even more remarkable results came when we tried to couple 3-dimension tensor
and scalar fields (with $\omega $ $\neq $ 0). The equations satisfied by the
matter variables themselves are formally the same as in standard Jordan,
Brans-Dicke theory. This was done in such a way, however, as to keep the
Lagrangian and action of matter itself unchanged. It is interesting to
investigate whether there exists another relativistic gravity theory which
does have the Newtonian theory as its weak-field limit. Consider a
3-dimensional analog of Jordan, Brans-Dicke theory where the temporal scalar
field will introduce some dynamics. A model based on this assumption does
declare that scalar curvature in empty space is not vanishes. Field
equations in a static case are determined by the 3-dimensional tensor field
(Riemann metric) and temporal scalar field.

\begin{eqnarray}
\begin{array}{c}
R_{\mu \nu }-\frac 12Rg_{\mu \nu }=\frac{8\pi }\phi T_{\mu \nu }-\frac
\omega {\phi ^2}\left( \phi _{,\mu }\phi _{,\nu }-\frac 12g_{\mu \nu }\phi
_{,\lambda }\phi ^{,\lambda }\right) -\frac 1\phi \left( \phi _{,\mu ;\nu
}-g_{\mu \nu }\Box \phi \right) , \\ 
\\ 
\Box \phi =-\frac{8\pi }{3+2\omega }T.
\end{array}
\label{12.2}
\end{eqnarray}

A simple calculation shows that the solution g$_{\mu \nu }$ = $\eta _{\mu
\nu }$ and $\phi $ = const completely satisfy these equations when
energy-momentum tensor of matter vanish. The most appropriate coordinates in
which to study the static spherically symmetric field are isotropic. Hence,
the metric will be assumed to have the form

\begin{eqnarray}
ds^2=e^{2\beta }\left( dr^2+r^2\left( d\theta ^2+\sin ^2\theta \,d\varphi
^2\right) \right) ,  \label{2.3}
\end{eqnarray}

where, $\beta $, $\phi $ and all matter variables will be assumed to be
function of r only. Expressing the line element in isotropic form gives:

\begin{eqnarray}
\begin{array}{c}
\frac{\beta ^{\prime }\phi ^{\prime }}\phi -\frac{2\beta ^{\prime }}r-\omega
\left( \frac{\,\phi ^{\prime \,}}\phi \right) ^2-2\beta ^{\prime \prime }-%
\frac{\phi ^{\prime \prime }}\phi =0 \\ 
\\ 
3\beta ^{\prime }+r\,\beta ^{\prime }\,^2+r\,\beta ^{\prime \prime }+\frac{%
\left( 1+r\,\beta ^{\prime }\right) \phi ^{\prime }}\phi =0 \\ 
\\ 
\frac 2r+\beta ^{\prime }+\frac{\phi ^{\prime \prime }}{\phi ^{\prime }}=0
\end{array}
\label{2.4}
\end{eqnarray}

One can obtain the exact vacuum solutions to these equations. There are a
various branches for this solutions corresponding to different ranges of the
arbitrary constants available. In the following these will be denoted by 
{\it k, m, h }and{\it \ q} have values in the indicated ranges. For example:

\begin{eqnarray}
\begin{array}{c}
\phi =\left( \frac{1-\frac mr}{1+\frac mr}\right) ^k, \\ 
\\ 
e^{2\beta }=\frac h{r^4}(1-\frac rm)^{-2k}\,m^{-2k}\left( r+m\right)
^{2k}\left( r^2-m^2\right) ^2.
\end{array}
\label{2.9}
\end{eqnarray}

and

\begin{eqnarray}
\begin{array}{c}
\phi =h\left( \frac{\frac 1r-4q\sqrt{2+\omega }}{\frac 1r+4q\sqrt{2+\omega }}%
\right) ^{2k}, \\ 
\\ 
e^{2\beta }=\frac m{r^4}\left( \frac{\frac 1r-4q\sqrt{2+\omega }}{\frac 1r+4q%
\sqrt{2+\omega }}\right) ^{2k}\left( \frac 1{q^2}-16r^2\left( 2+\omega
\right) \right) ^2.
\end{array}
\label{2.9}
\end{eqnarray}

where {\it k} = $\pm \frac{\sqrt{2}}{\sqrt{2+\omega }}.\,$Or

\begin{eqnarray}
\begin{array}{c}
\phi =h\left( \frac{m+r}{m-r}\right) ^k, \\ 
\\ 
e^{2\beta }=\frac{\left( m-r\right) ^4}{8m^2r^4h^2}\left( \frac{m+r}{m-r}%
\right) ^{2-2k}\left( 2+\omega \right) .
\end{array}
\label{2.9}
\end{eqnarray}

\section{Conclusion}

The objectives of this paper are more limited then formulation of a complete
theory of time as a scalar field on the manifold. Such a program would
consist of the formulation of a suitable field theory and boundary and
initial value conditions. The theory to be developed represents a
modification of general relativity can be interpreted as a dynamical theory
of evolution of three dimensional Riemannian geometry. Any mechanism that
allows a bridge to be established between such theories and the ordinary 4
dimensional gravity is of vital interest. One possible way to proceed is to
employ the equivalence between the (3 + 1) Einstein theories with 3
dimensional analog of Jordan, Brans-Dicke theories (with $\omega $ = 0). The
dynamics induced by the new temporal field contains all features that one
expects in a theory of gravitation. These ideas may be continued by
suggested considering the more general case $\omega $ $\neq $ 0 leading to
the theory with an additional temporal scalar field. It is not a completely
geometrical theory of gravitation, as gravitational effects are described by
a scalar field in a Rimannian manifold. Thus, gravitational effects are in
part geometrical and in part due to scalar interaction. There is formal
connection between this theory and that of Jordan, Brans Dicke, but there
are differences and the physical interpretation is quite different.

We expect very interesting consequences of dynamics from these
considerations, which will be investigated in future work.

\vspace*{1.5cm}

{\bf Acknowledgements.} The author is grateful to dr. Rolando Cardenas for
the kind invitation to make a report at the Santa Clara 2004. International
Workshop on Gravitation and Cosmology (31 may-3 June, 2004 Central
University of Las Villas, Cuba).

\end{document}